\documentclass{jfm}

\usepackage{graphicx}
\usepackage{newtxtext}
\usepackage{newtxmath}
\usepackage{natbib}
\usepackage{hyperref}
\usepackage[usenames,dvipsnames]{xcolor}
\hypersetup{
    colorlinks = true,
    urlcolor   = blue,
    citecolor  = black,
}

\newcommand{\RomanNumeralCaps}[1]
\linenumbers

\title{Slumping regime in lock-release turbidity currents}

\author{Cyril Gadal\aff{1}
  \corresp{\email{cyril.gadal@imft.fr}},
  M. J. Mercier\aff{1},
  M. Rastello\aff{2}
 \and L. Lacaze\aff{1}}

\affiliation{\aff{1}Institut de M\'ecanique des Fluides de Toulouse (IMFT), Universit\'e de Toulouse, CNRS, Toulouse, France
\aff{2}Univ. Grenoble Alpes, CNRS, Grenoble INP, LEGI, 38000 Grenoble, France}

\begin{document}
\maketitle

\textcolor{blue}{An edited version of this draft was published by Cambridge University Press:
\begin{itemize}
  \item Gadal, C., Mercier, M., Rastello, M., \& Lacaze, L. (2023). Slumping regime in lock-release turbidity currents. Journal of Fluid Mechanics, 974, A4. doi:10.1017/jfm.2023.762  
\end{itemize}
}

\qquad

\begin{abstract}
  Most gravitational currents occur on sloping topographies, often in the presence of particles that can settle during the current propagation. Yet an exhaustive exploration of associated parameters in experimental studies is still lacking. Here, we present an extensive experimental investigation of the slumping regime of turbidity (particle-laden) currents in two lock-release (dam-break) systems with inclined bottoms. We identify three regimes controlled by the ratio between settling and current inertia. (i) For negligible settling, the turbidity current morphodynamics corresponds to that of saline homogeneous gravity currents, in terms of velocity, slumping (constant-velocity) regime duration and current morphology. (ii) For intermediate settling, the slumping regime duration decreases to become fully controlled by a particle settling characteristic time. (iii) When settling overcomes the current initial inertia, the slumping (constant-velocity) regime is not longer detected. In the first two regimes, the current velocity increases with the bottom slope, of approximately $35~\%$ between $0^\circ$ and $15^\circ$. Finally, our experiments show that the current propagates during the slumping regime with the same shape in the frame of the moving front. Strikingly, the current head is found to be independent of all experimental parameters covered in the present study. We also quantify water entrainment coefficients $E$ and compare them with previous literature, hence finding that $E$ increases rather linearly with the current Reynolds number.
\end{abstract}


\section{Introduction}
\label{sec:intro}

Turbidity currents are gravity-driven flows induced by the presence of suspended particles, in addition to other processes that may affect the density, such as temperature, salinity or humidity. They occur ubiquitously in nature, from submarine turbidites to powder snow avalanches and volcanic pyroclastic flows, and are almost always sources of potential natural hazards \citep[e.g.][]{dobran1994, stethem2003, carter2014, clare2020}.

These currents have been studied extensively along with homogeneous (saline) density-driven gravity currents for almost a century, by means of experiments \citep[e.g.][]{simpson1980, Rastello2002, Dai2014, Lippert2020}, theoretical analyses \citep[e.g.][]{benjamin1968, huppert1998, Hogg2001, ungarish2009book} and numerical simulations \citep[e.g.][]{necker2002high, blanchette2005, Cantero2007, Cantero2012a, Ottolenghi2016}. Among these studies, a major source of interest has been to predict the front velocity $u_{c}$ of the current. Dimensionally, a current of height $h$ and density $\rho_{0}$, hence of density difference $\Delta\rho$ with respect to the ambient density $\rho_{\rm f}$, would have a front velocity scaling as
\begin{equation}
  \label{eq:scaling_velocity}
  u_{c} \propto \sqrt{g \frac{\Delta\rho}{\rho_{0}} h}.
\end{equation}
Many works have been devoted to capturing the exact value of the proportionality factor. The pioneering work of \citet{von1940} leads to $\sqrt{2}$ in the case of steady unbounded flows, further extended to account for finite flow depth \citep{benjamin1968, Rottman1983, Ungarish2005}, energy conservation/dissipation \citep{shin2004, borden2013} or non-Boussinesq density difference \citep{ungarish2007, ungarish2011, konopliv2016}.

Gravity currents can be generated by a constant source of buoyancy \citep[e.g.][]{Britter1980, Baines2001, Cenedese2008, Lippert2020} or can result from the instantaneous release of a limited volume of buoyant fluid. In the latter case, dam-break (or lock-exchange) systems are a common set-up to study the features of the resulting currents \citep[e.g.][]{simpson1972, huppert1980, Rottman1983, bonnecaze1993, Ungarish2005, ungarish2007, ungarish2011, Chowdhury2011, khodkar2017, balasubramanian2018, maggi2022}. The heavier (or lighter) fluid is kept separated from the ambient by a locked gate, which is opened suddenly to generate the current. For high Reynolds number flows, the front velocity of the resulting currents can evolve through different regimes \citep{huppert1980}. After a short transient acceleration stage~\citep{Cantero2007}, first there is a regime of constant velocity, called the slumping regime, as the current gains inertia thanks to the collapse of the heavy (or light) fluid column, which lasts about 5 -- 15 lock lengths depending on the geometry \citep{Rottman1983, Ungarish2005, ungarish2009book}. If inertia dominates the flow, then the receding rarefaction wave during column slumping hits the back wall and reflects towards the current nose, modifying its velocity into an inertial regime, where the front position evolves as $t^{2/3}$. The current eventually enters regimes dominated by either viscosity \citep{huppert1980}, friction and entrainment \citep{bonnecaze1999, Hogg2001} or particle settling velocity \citep{bonnecaze1993, bonnecaze1995, hallworth1998, huppert1998, hogg2000, harris2001}.

In lock-release systems, these previous studies have focused on the impact of the settling velocity $u_{\rm s}$ only on the long-term dynamics of the current, typically during the inertial regime, and after. This corresponds to small values of the settling number $u_{\rm s}/u_{\rm c}$, for which particle settling does not impact the initial slumping regime. Likewise, theoretical studies have also restrained to asymptotically small settling number values that allow for analytical development in depth-averaged models~\citep{hogg2000, harris2001}. However, recently the study of \citet{ikeda2021} observed qualitatively increasing deviations of particle-laden from saline currents in all propagation regimes as the settling number increases. Literature on constant inflow turbidity currents has also observed similar differences, especially concerning the volume occupied by the current \citep{bonnecaze1999, Lippert2020, Wells2021}. One purpose of this study is therefore to quantify the dynamics of constant-volume turbidity currents from low to strong settling, across a wide range of settling number values.

Lock-release homogeneous and turbidity gravity currents on an inclined plane, which induces an extra-driving force due to the weight of the current, have also been studied in the literature~\citep{Beghin1981, Rastello2002, Seon2005, Birman2007, Maxworthy2007, Dai2013, Dai2014, Steenhauer2017, Xie2023}. Hence, after the slumping regime, the current dynamics is characterized at intermediate times by an acceleration phase, later followed by a deceleration resulting from buoyancy loss induced by water entrainment, increasingly important at large slopes. Importantly, for saline homogeneous currents, the Navier-Stokes simulations of \citet{Birman2007} reported that the initial constant-velocity (slumping) regime still exists at early times, during which the slope-induced acceleration is negligible. They also reported an increase in slumping velocity with the bottom slope, of approximately $15~\%$ between $0^\circ$ and $15^\circ$. However, the experiments of \citet{Maxworthy2007} measured much larger variations, up to $30~\%$. Note that, to the authors' knowledge, no similar study is available in the literature concerning turbidity currents. Therefore, a second purpose of this study is to quantify experimentally the impact of an inclined bottom on the slumping regime dynamics of particle-laden currents.

In this work, we present lock-release experiments of turbidity currents, where we vary systematically the initial volume fraction, the bottom slope and the particle diameter (and thus the settling velocity), hence extending previous works to a larger range of these control parameters in two experimental devices. We focus particularly on the slumping regime, for which we map its existence and quantify its duration as well as the related current morphodynamics (velocity, shape) and water entrainment. In the paper, we also focus on rationalizing existing results with those obtained in the present study into a relevant parameter map characterizing the flow regimes.

\section{Methods}

\subsection{Experimental set-ups}
\label{sec:set_up}
\begin{figure}
  \includegraphics[scale=1]{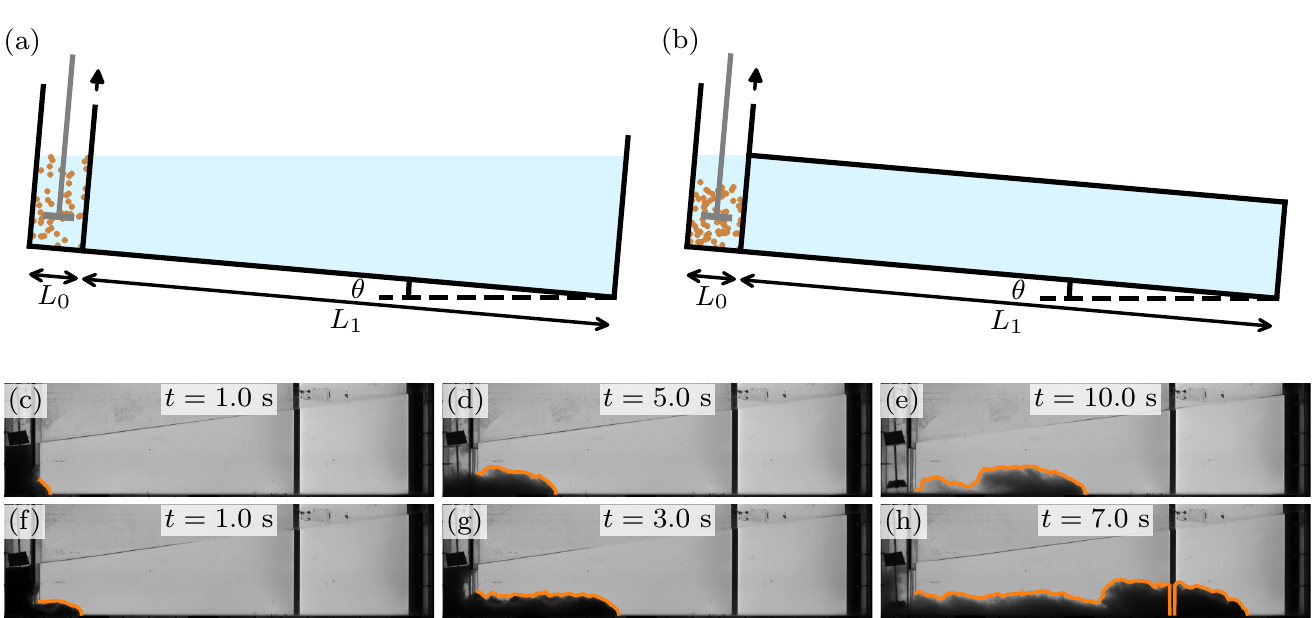}
  \caption{(a,b) Sketches of the experimental set-ups 1 et 2, respectively. (c--h) Snapshots of experiments using the silica sand ($d \sim 120~\mu\textup{m}$, $u_{\rm s} = 0.74~\textrm{cm}~\textrm{s}^{-1}$), $\theta=7.2^\circ$ and for initial volume fractions of (c--e) $\phi=0.87~\%$ and (f--h) $\phi=6.4~\%$ . The orange lines show the extracted current contours.}
  \label{fig:fig1}
\end{figure}

In this study, most of the experiments are done using the dam-break experimental set-up sketched in figure~\ref{fig:fig1}(a), later referred to as set-up 1. The tank, $150$~cm long ($L_{0} + L_{1}$) and $20$~cm wide ($W_{0}$), is filled with water, and divided into two parts by a sluice gate at $10~\textup{cm}$ ($L_{0}$) from the left-hand of the tank. It forms a reservoir on the left-hand of the tank in which we prepare an initial volume of particle suspension $V_{0} \simeq 3.9$~l by strongly stirring a known mass of particles $m_{0}$ within the water. Finally, the tank is inclinable at various angles up to $7^\circ$, and we keep the water height at the gate position constant, equal to $20~\textup{cm}$. The resulting variation of the initial volume $V_{0}$ is accounted for, however small compared to the experimental uncertainties.
At the beginning of the experiments, as soon as the stirring is stopped (after less than $0.5~\textup{s}$), the sluice gate is opened manually almost entirely, up to $\approx 1~\textup{cm}$ below the water surface to limit as much as possible the generation of surface waves. The slumping of the column and the resulting turbidity current are followed by a camera while using a backlight as a light source (see figures~\ref{fig:fig1}(c--h)).

In order to explore further the influence of the bottom inclination, another experimental set-up is used (set-up 2, see figure~\ref{fig:fig1}(b)). Here, the tank can be further inclined thanks to the presence of a rigid lid covering the water surface, keeping the water height to $50~\textrm{cm}$. Here, $L_{0} = 10~\textrm{cm}$, $L_{1} = 340~\textrm{cm}$ and $W_{0} = 10~\textrm{cm}$. Note that in this experimental setup, the suspension is filling not the entire reservoir height, but approximately $50~\%$--$75~\%$ of that height. Nevertheless, the suspension is checked qualitatively by light attenuation to be homogeneously suspended up to its maximum height, and the associated initial volume of suspension $V_{0}$ is extracted from images prior to opening the gate. Finally, in this set-up, the current is illuminated from the top, and not using backlighting.

\subsection{Parameter space and relevant dimensionless quantities}
Most experiments are done with silica sand grains of diameter $d \simeq 120~\mu\textup{m}$. For these particles, the tank inclination $\theta$ is varied from $0^\circ$ to $7^\circ$ in set-up 1, and from $7^\circ$ to $15^\circ$ in set-up 2. Then, in set-up 1 and for $\theta = 7^\circ$, the particle settling velocity $u_{\rm s}$ is varied by using glass beads (\copyright Silibeads) of mean diameter ranging from $60~\mu$m to $250~\mu$m, corresponding to $u_{\rm s} \in [0.3,\, 3.2]~\textrm{cm}~\textrm{s}^{-1}$. As particles are either glass beads or silica sand, we take for all cases $\rho_{\rm p} = 2.65~\textrm{g}~\textrm{cm}^{-3}$. The particle properties are detailed in Appendix~\ref{sec:grain_properties}.
For all bottom slopes and settling velocities, the initial volume fraction is varied in the range $\phi \in [0.25,\, 30]~\%$. The corresponding excess density of the fluid/particle suspension with respect to the ambient water, $\Delta\rho = \rho_{0} - \rho_{\rm f}$ (where $\rho_{0} = \rho_{\rm f} + (\rho_{\rm p} - \rho_{\rm f})\phi$), therefore varied between
$3~\textrm{kg}~\textrm{m}^{-3}$ and $600~\textrm{kg}~\textrm{m}^{-3}$. Finally, we also perform experiments with homogeneous saline (without particles) gravity currents in set-up 1, in order to make a direct comparison between turbidity and homogeneous currents. For that purpose, the density in the reservoir is varied from $1002~\textrm{kg}~\textrm{m}^{-3}$ to $1250~\textrm{kg}~\textrm{m}^{-3}$ to explore the same range of $\Delta\rho$ values. This results in a total of 169 experimental runs.

Each experiment is characterized by three initial quantities, the bottom slope $\theta$, the volume fraction $\phi$ or equivalently the excess density $\Delta\rho$ as will be discussed later, and the particle settling velocity $u_{\rm s}$. For saline homogeneous cases, only slope $\theta$ and excess density $\Delta\rho$ then characterize the system. Note that the initial aspect ratio of the reservoir, $a = h_{0}/L_{0}$, is kept nearly constant in each set-up, equal to 2 in set-up 1, and $\simeq 3$ in set-up 2. Its influence will be discussed in the paper.
Following the available literature, we define velocity and time scales as
\begin{equation}
  u_{0} = \sqrt{g'h_{0}}
\end{equation}
and
\begin{equation}
  t_{0} = \frac{L_{0}}{u_{0}},
\end{equation}
where $h_{0} = V_{0}/(L_{0} W_{0})$ is the average initial heavy fluid height, and $g' = g\Delta\rho/\rho_{\rm f}$ is the reduced gravity. In the case of turbidity currents, we also write $g' = g(\rho_{\rm p} - \rho_{\rm f})\phi/\rho_{\rm f}$, where $\rho_{\rm p}$ and $\rho_{\rm f}$ are the particle and fluid densities.
This velocity scale can be used to define a Reynolds number $\mathcal{R}e$ and a Stokes number $\mathcal{S}$ as the control dimensionless parameters based on the initial conditions
\begin{equation}
  \mathcal{R}e = \frac{u_{0} h_{0}}{\nu}, \quad \mathcal{S} = \frac{L_{0}}{u_{0}}\frac{u_{\rm s}}{h_{0}},
\end{equation}
where $\nu$ is the water kinematic viscosity. Here, $\mathcal{S}$ is a ratio of time scales, thus depending on the reservoir aspect ratio $a$ and the dimensionless settling velocity $u_{\rm s}/u_{0}$, also called the settling number in previous studies~\citep[e.g.][]{Lippert2020}. In our experiments, we then have $\mathcal{R}e \in [2 {\times} 10^{4},\, 4 {\times} 10^{5}]$, and $\mathcal{S} \in [0.002,\, 0.05]$. Note that the Rouse number $u_{\rm s}/u_{*}$, where $u_{*}$ is a shear velocity, has also been used instead of $\mathcal{S}$ to quantify the ability of the particles to remain in suspension~\citep{Wells2021}. However, it requires local measurements of the velocity fluctuations, which we do not perform here.

Importantly, in lock-release systems, $u_{0}$ is the only velocity scale associated with the gravity current, such that the initial Froude number reduces to unity for all experiments. On the other hand, we define a Froude number as the dimensionless current velocity in the slumping regime:
\begin{equation}
  \mathcal{F}r = \frac{u_{\rm c}}{u_{0}},
\end{equation}
where $u_{\rm c}$ is the current velocity in the slumping (constant-velocity) regime.

\section{Current dynamics during the slumping regime}
\label{sec:constant_vel_regime}

In this section, we focus on the current dynamics and shape during the slumping regime, and explore the effect of the bottom slope and particle settling velocity.

\subsection{Nose position and velocity}
\label{sec:nose_position}

\begin{figure}
  \includegraphics[scale=1]{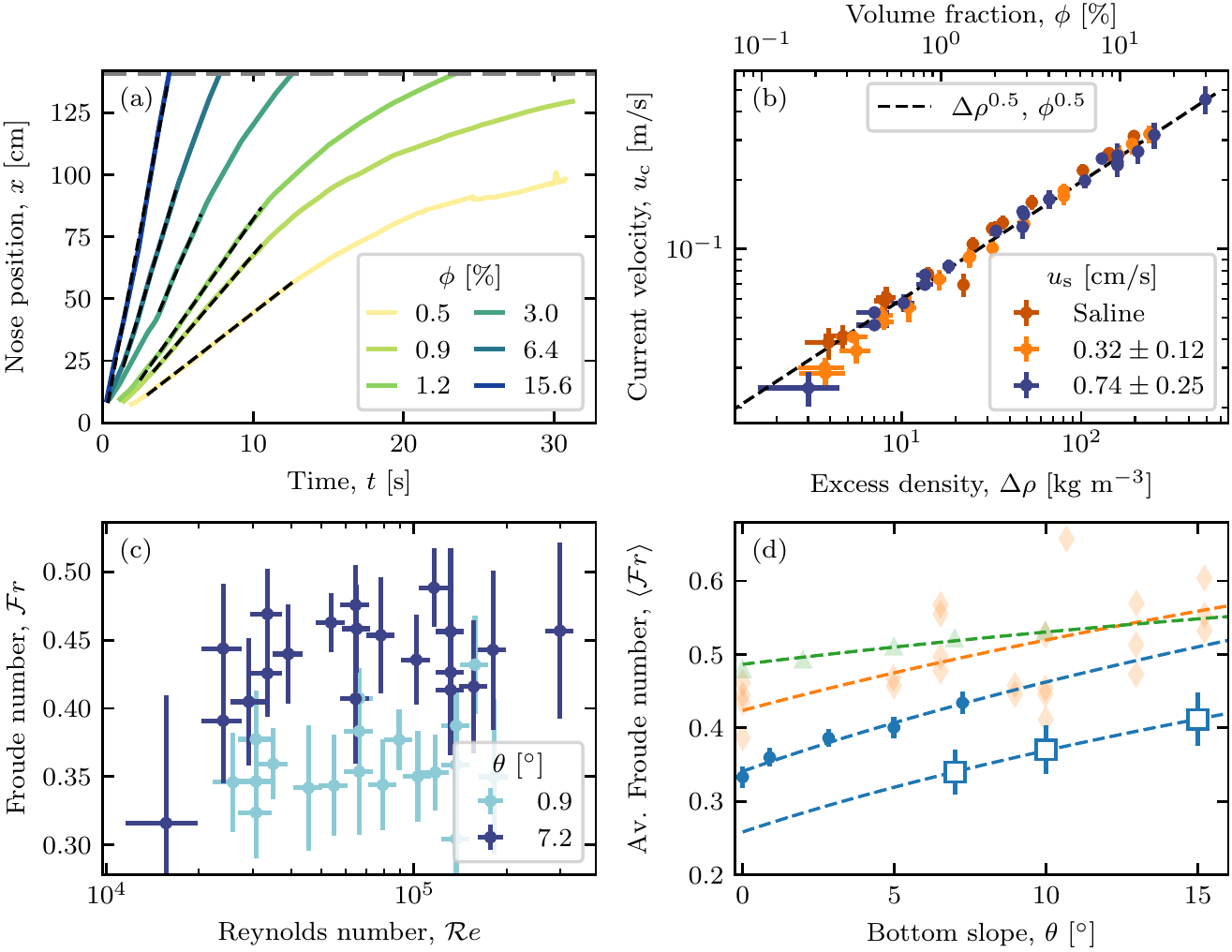}
  \caption{(a)~Current nose position as a function of time for various initial volume fractions and for a bottom slope $\theta=7^\circ$ and $u_{\rm s} = 0.74~\textrm{cm}~\textrm{s}^{-1}$ (for clarity purposes, not all experiments are shown here). The black dashed lines are linear fits on the constant-velocity regime, whose slopes give the current velocity $u_{\rm c}$. The grey dashed line indicates the end of the tank. (b)~Current velocity $u_{\rm c}$ as a function of the excess density and the volume fraction, for a bottom slope $\theta=7^\circ$ and different settling velocities. (c)~Current Froude number as a function of the initial Reynolds number for two different bottom slopes for $u_{\rm s} = 0.74~\textrm{cm}~\textrm{s}^{-1}$. (d) Current Froude number averaged over the initial volume fraction as a function of the bottom slope for $u_{\rm s} = 0.74~\textrm{cm}~\textrm{s}^{-1}$. Circles correspond to set-up 1, and empty squares to set-up 2. Orange diamonds correspond to the experiments of \citet{Maxworthy2007}, and green triangles to the numerical simulations ($\mathcal{R}e = 4000$) of \citet{Birman2007}, both for homogeneous saline currents. The dashed lines are fits of \eqref{eq:Fr_slope} to the four previous datasets, leading to $(Fr_{0}, C)$ equal to $(0.34, 5)$, $(0.26, 6)$, $(0.42, 3)$ and $(0.48, 1)$, respectively.}
  \label{fig:fig2}
\end{figure}

\subsubsection{Slumping behaviour}
\label{sec:slumping_behaviour}

First, we start by tracking the current front position, displayed as a function of time in figure~\ref{fig:fig2}(a) for different initial volume fractions $\phi$ (or equivalently, different $\Delta\rho$) and $\theta=7^{\circ}$. After a short acceleration phase corresponding to the early collapse of the heavy fluid column dominated by vertical motion (for details, see \citet{Cantero2007}), all experiments exhibit a regime where the current propagates at a constant front velocity $u_{c}$ (dashed black lines in figure~\ref{fig:fig2}(a)), also known as the slumping regime.

In this regime, the measured velocity comes from the lossless conversion of the initial potential energy of the heavy fluid column, $\Delta\rho g h_{0}$, into kinetic energy, $(1/2)\rho_{0}u_{\rm c}^{2}$, leading to
\begin{equation}
  \label{eq:vcourant_1}
  u_{\rm c} \propto \sqrt{\frac{\Delta\rho}{\rho_{0}}gh_{0}},
\end{equation}
where $\rho_{0}$ is the initial heavy fluid density. The prefactor of \eqref{eq:vcourant_1} is notably proportional to $\sqrt{\rho_{0}/\rho_{f}}$ \citep{von1940, benjamin1968, shin2004}, leading to
\begin{equation}
  \label{eq:vcourant_2}
  u_{\rm c} \propto u_{0}
\end{equation}
As shown in figure~\ref{fig:fig2}(b), the current velocity indeed scales as $\Delta\rho^{1/2}$ (or equivalently, $\phi^{1/2}$), as expected from \eqref{eq:vcourant_2}. This also corresponds to a constant Froude number $\mathcal{F}r=u_{\rm c}/u_{0}$ as shown in figure~\ref{fig:fig2}(c) (dark blue symbols for $\theta=7^{\circ}$).
Varying the particle settling velocity while keeping the bottom slope at $7^\circ$ impacts neither the scaling of \eqref{eq:vcourant_2} nor its prefactor (see figure~\ref{fig:fig2}(b)), which remains within $20~\%$ of the one corresponding to homogeneous saline gravity currents (red symbols).

To conclude, on an inclined bottom, the first propagation regime of our turbidity currents is characterized by a time-independent velocity, similar to the behaviour of a gravity current on a horizontal bottom. Surprisingly, the corresponding velocity values are nearly independent of $ \mathcal{R}e$ and $\mathcal{S}$ in the large ranges considered here.

\subsubsection{Effect of the bottom slope}

On the other hand, changing the bottom slope more clearly affects the current velocity. As shown in figure~\ref{fig:fig2}(c), currents on slope of $7^{\circ}$ are nearly $30\%$ faster than those on $\theta = 1^\circ$. After averaging across the initial volume fraction $\phi$ (in which no clear dependency is observed, as explained previously), we find that the Froude number $\langle \mathcal{F}r \rangle$ increases rather linearly with the bottom slope $\theta$ (see figure~\ref{fig:fig2}(d)). The slope of this linear relationship is recovered in set-up 2 and in the experiments of \citet{Maxworthy2007} for saline currents, but not in the Navier-Stokes numerical simulations of \citet{Birman2007}, who report a much weaker increasing trend.

This increase in the Froude number with the bottom slope is not necessarily obvious. On one hand, on a flat bottom, the constant velocity in the slumping regime results from a balance between inertia and pressure gradient. On the other hand, the slope adds a constant forcing term that could result in an accelerated flow. In this case, the full balance of these different terms does not then lead to a constant velocity. Yet, it is observed clearly as a constant in our experiments, as well as in previous studies in the literature~\citep{Birman2007, blanchette2005, Xie2023}.
Following \citet{ross2002study} and \citet{Birman2007}, the along-slope component of gravity, $g'\sin\theta$, needs a dimensionless time of $O(1/\sin\theta)$ to accelerate the flow at a velocity of $O(u_{0})$. Hence on small slopes, this does not modify the slumping equilibrium, and the observed constant-velocity regime is still attributed here to a slumping regime including inertia and pressure gradient, and not to a frictional-buoyancy equilibrium that would be obtained at larger slopes and/or longer times~\citep{Britter1980}.

Yet this does not explain the dependence of the slumping velocity on the bottom slope. This variation could instead result from the early transient acceleration phase, during which the current accelerates from 0 to the constant slumping velocity. The complicated dynamics of this transient phase involves significant vertical motions, and significant interfacial friction responsible for the development of vortices~\citep{Cantero2007}. We then consider an energetic balance between the initial state and the end of this early acceleration phase:
\begin{equation}
  \left[\frac{1}{2}\rho_{0}u_{\rm c}^{2} - B \Delta\rho g \sin\theta L\right] - A \Delta\rho g \cos\theta h_{0} =  - \frac{1}{2}C_{\rm d}\rho_{0}u_{\rm c}^{2} \frac{L}{h_{0}},
\end{equation}
where $A$, $B$ and $C_{\rm d}$ are constants accounting for details of the reservoir evolution during the transient phase. Note that, depending on the lock geometry, they might induce a second-order dependency on the bottom slope $\theta$, which we will neglect here.
Here, $L$ accounts for the along-slope distance over which the current moves during this phase, which \citet{Cantero2007} has observed to be independent of the lock aspect ratio or initial buoyancy, $L = 0.3 h_{0}$. As a result, we use here $L = D h_{0}$ with $D$ constant, possibly neglecting second-order dependencies on $\theta$.
Hence, under the Boussinesq approximation,
\begin{equation}
  \label{eq:Fr_slope}
  \mathcal{F}r(\theta) \equiv \frac{u_{\rm c}}{u_{0}} = \sqrt{\frac{2 A}{1 + D C_{\rm d}}} \sqrt{\cos\theta + \frac{D B}{A}\sin\theta}\equiv \mathcal{F}r_0 \sqrt{\cos\theta + C\sin\theta},
\end{equation}
where $C = D(B/A)$ and $\mathcal{F}r_0 = \mathcal{F}r(\theta = 0)$. For small slopes ($\theta \to 0^\circ$), \eqref{eq:Fr_slope} can be approximated by a linear relationship in $\theta$, as suggested previously.

The fits of \eqref{eq:Fr_slope} to four datasets are shown in figure~\ref{fig:fig2}(d). Although \eqref{eq:Fr_slope} is able to represent the data well, the fits are poorly constrained due to the small number of experimental points, or the large dispersion in the dataset of \citet{Maxworthy2007}. This is especially true for the parameter $C$, whose uncertainty can reach $100~\%$ ($95~\%$ confidence interval). It is, however, found to be similar in all experimental datasets, but smaller for the numerical simulations of~\citep{Birman2007}. Note that the order of magnitude of $C$ implies that the linearized version of \eqref{eq:Fr_slope} could be used up to $\theta \sim 1^\circ$, hence justifying here the use of the nonlinearized form of \eqref{eq:Fr_slope}.
The resulting values of the Froude number for $\theta = 0^\circ$, much better constrained, are different across datasets and generally smaller than $\mathcal{F}r = 0.5$ predicted on a non-inclined tank by the simple steady model of \citet{benjamin1968}. Depth-averaged models, including the properties of dam-break configurations \citep{Ungarish2005}, with non-Boussinesq effects~\citep{ungarish2007} or the motion of the lighter fluid in the upper layer~\citep{ungarish2011}, also lead to larger predicted Froude numbers. Previous measurements on saline and turbidity currents have reported Froude numbers in better agreement with these theories \citep[e.g.][]{shin2004, lowe2005, nogueira2014, sher2015}, but also smaller ones in agreement with those measured in this study \citep{Longo2018, balasubramanian2018}.

The general discrepancies between the different datasets could come from geometrical differences between the corresponding experimental set-ups. In the experimental set-up of \citet{Maxworthy2007}, the heavy fluid is released way below the water surface (partial depth release), whereas only full-depth releases were performed in set-up 1. According to the predictions of \citet{Ungarish2005}, this can lead to an increase of the velocity of almost $50~\%$, which matches well the discrepancy between the two studies. Note, however, that the simulations of \citet{Birman2007}, leading to the highest $\mathcal{F}r(0)$ values, are also full-depth releases.
In set-up 2, the ratio between typical current heights and the tank width is approximately 1.5, compared to 0.5 in set-up 1. Therefore, we can also expect energy dissipation induced by friction at the walls of the tank to be much larger, hence explaining the lower measured Froude numbers.
Finally, the different set-ups also have different lock aspect ratios $a = h_{0}/L_{0}$. More importantly, set-ups of \citet{Maxworthy2007} and \citet{Birman2007} are in lock-exchange configurations with $a < 1$ ($a = 0.5$ and $a = 0.1$, respectively), while our experiments always have $a > 1$ ($a = 2$ and $a \simeq 3$ in set-ups 1 and 2, respectively). We find a linear decrease of the Froude number with $a$ (not shown here), a trend already reported by \citet{bonometti2011numerical}, although weaker ($a^{1/4}$). Dedicated experiments would be required to study further in detail the slumping of the column, and its dependence on the geometry of the experiments, by changing, for instance, the lock aspect ratio $a$.

To conclude, the conceptual model \eqref{eq:Fr_slope} reproduces the $\theta$ dependency of the Froude number for both saline and turbidity currents obtained in the present configuration, extending available results from the literature focusing mostly on the zero-slope configuration to predict the front velocity of the current. However, such an approach does not allow the influence of the particle settling to be isolated compared to the situation of a homogeneous saline current. This will be discussed in the following.

\subsection{Existence and duration of the slumping regime}
\label{sec:settling_velocity}

The previous section discussed the value of the current-velocity during the constant velocity (slumping) regime. However, the latter could not be detected in every experiment. In the following, we discuss its duration and existence with respect to the settling velocity $u_{\rm s}$ and the excess density $\Delta\rho$, while keeping the bottom slope at $7^\circ$.

\subsubsection{Existence of the constant-velocity regime}
\label{sec:existence_regime}

\begin{figure}
  \includegraphics[scale=1]{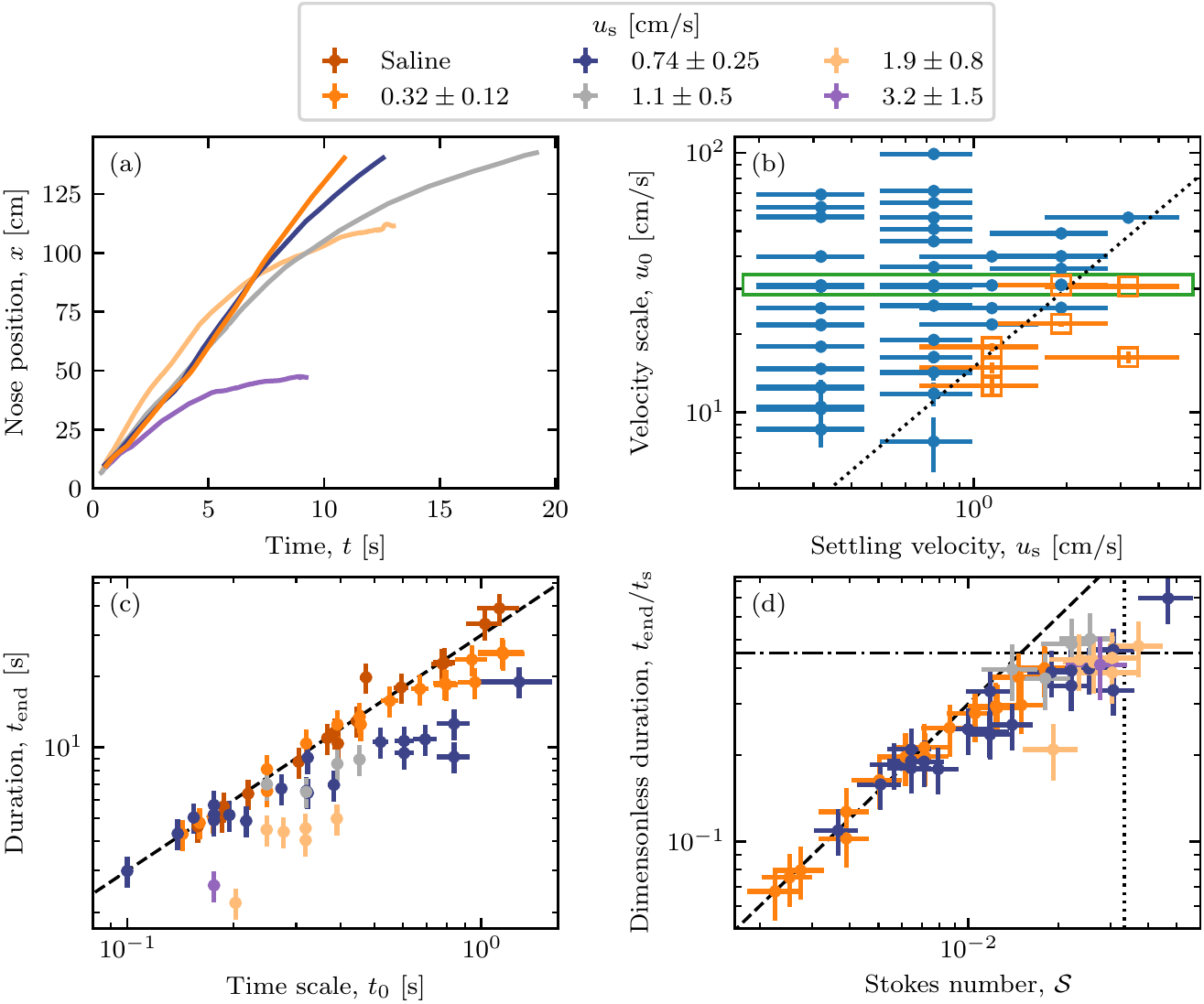}
  \caption{(a) Current nose position as a function of time for various particle settling velocities, at a fixed volume fraction $\phi = 3~\%$. (b) Regime diagram indicating currents for which the constant-velocity regime is detected (blue dots) and those where it is not (orange squares). See section~\ref{sec:existence_regime} for more details. The black dotted line indicates a possible linear regime separation, corresponding to $u_{\rm s}/u_{0} \simeq 0.067$. The green rectangle indicates experiments displayed in (a). (c) Duration of the constant-velocity regime as a function of the current characteristic time scale, for various particle settling velocities. (d) Same as (c), but with rescaling both axes by a settling time scale $t_{\rm s} = h_{0}/u_{\rm s}$. In both plots, the black dashed line indicates $t_{\rm end} = 30 t_{0}$. In (d), the vertical dotted line indicates the limit $\mathcal{S} \simeq 0.033$ (see (b) and section~\ref{sec:settling_velocity}), and the horizontal dash-dotted line indicates the limit $t_{\rm end} = 0.45t_{\rm s}$. In this figure, experiments were performed in set-up 1 ($a = 2$) for $\theta = 7^\circ$.}
  \label{fig:fig3}
\end{figure}

Figure~\ref{fig:fig3}(a) shows the influence of the settling velocity on the nose propagation of currents at a fixed $\Delta\rho = 45~\textrm{g}~\textrm{cm}^{-3}$ ($\phi = 3~\%$). As discussed in section~\ref{sec:nose_position}, all curves exhibit the same initial constant velocity at a given slope $\theta$, except the largest settling velocity, for which no clear constant-velocity regime can be observed.
For all our experiments, we classify the cases with (blue dots) or without (orange squares) a constant-velocity regime in figure~\ref{fig:fig3}(b), in a ($u_{\rm s}$,$u_0$) diagram. The cases shown in Figure~\ref{fig:fig3}(a) are indicated by a horizontal green rectangle. It highlights a sharp transition at a dimensionless settling velocity $u_{\rm s}/u_{0} \simeq 0.067$ (black dotted line in figure~\ref{fig:fig3}(b)) separating currents for a constant-velocity regime is observed, with a velocity $\mathcal{F}r(\theta)$ independent of $\mathcal{S}$, from currents for which the velocity always decreases.

\subsubsection{Duration of the constant-velocity regime}

The duration of the constant-velocity regime, i.e. the time $t_{\rm end}$ at which the front evolution is observed to deviate from a linear trend with $t$, is shown to decrease as the settling velocity increases (see figure~\ref{fig:fig3}(a)). In the case of homogeneous saline gravity currents, this duration is approximately $t_{\rm end} \simeq 30 t_{0}$ as shown in figure~\ref{fig:fig3}(c). The latter result is in agreement with previous experiments and shallow-water modelling, and corresponds to the duration needed for the bore (current nose of the upper light fluid layer) to reach the nose of the heavy fluid current \citep{Rottman1983, Ungarish2005}. Note that previous studies have reported a prefactor of $t_{\rm end} \propto t_{0}$ between 20 and 30, which corresponds to travel distance 7--12 reservoir lengths, \citep{Rottman1983, Ungarish2005, Chowdhury2011, nogueira2014, sher2015, Ottolenghi2016}. The difference may result essentially from the difficulty in measuring $t_{\rm end}$ \citep{Rottman1983, ungarish2009book, Ottolenghi2016}. Then, in our experiments, slumping regime durations $t_{\rm end}$ shorter than $30~t_{0}$ can be attributed not to the transition to the classical inertial regime in $t^{2/3}$, but rather to a loss of buoyancy induced by another process, such as entrainment, or more probably particle settling, as discussed below.

For the smallest glass beads ($u_{\rm s} = 0.32~\textrm{cm}~\textrm{s}^{-1}$), figure~\ref{fig:fig3}(c) shows that they behave similarly to the saline gravity currents, except for slow currents (low volume fractions, high $t_{0} = L_{0}/u_{0}$) that exhibit smaller $t_{\rm end}$. As the settling velocity increases, an increasing number of cases do not follow this trend, more likely for large $t_{0}$ values, until all currents exhibit smaller $t_{\rm end}$ values.

By using $t_{\rm s} = h_{0}/u_{\rm s}$ as a characteristic settling time, which corresponds to the time required for a particle to settle over the initial column height, we obtain a good collapse of the data at various settling velocities (see figure~\ref{fig:fig3}(d)). The resulting trend, whose horizontal axis is now controlled by the Stokes number $\mathcal{S}$, exhibits a transition between two regimes. For small values of $\mathcal{S}$, the settling is negligible and $t_{\rm end}$ scales with $t_{0}$, as for saline density currents (black dashed line). For $\mathcal{S}$ larger than $0.01$, settling can no longer be neglected. The curve then transitions to a regime controlled entirely by particle settling, $t_{\rm end} \propto t_{\rm s}$ (dash-dotted line), or equivalently, $t_{\rm end}/t_{0} \propto \mathcal{S}^{-1}$. The trend stops at $\mathcal{S} \simeq 0.033$, the limit over which the constant-velocity regime is no longer observed.

The data presented in figure~\ref{fig:fig3} come from experiments performed in set-up 1, for which $a=2$ is kept constant. As such, it does not allow us to assess the relevance of $a$ (on which $S$ depends) on the control of the slumping regime duration.
However, by comparing data from set-ups 1 and 2 (different $a$) for the same particles (thus same $u_{\rm s}$), one can observe that a good collapse is obtained after rescaling by the settling time $h_{0}/u_{\rm s}$ (see Appendix figure~\ref{fig:figure_tend_slope}). This highlights the relevance of the lock aspect ratio $a$ in the control of the constant-velocity regime duration.
Finally, it has to be noted that no dependence of $t_{\rm end}$ with the bottom slope is found in the range of parameters covered here (see Appendix figure~\ref{fig:figure_tend_slope}). The latter result is not necessarily obvious, as the current velocity in the slumping regime was shown to depend on $\theta$ in section~\ref{sec:constant_vel_regime}. However, the slope is a second-order effect here, as previously explained.

In conclusion, the slumping regime is controlled dominantly by two parameters. On the one hand, its duration $t/t_{0}$ is obtained to depend mainly on $\mathcal{S}$, while on the other hand, the corresponding velocity $\mathcal{F}r = u_{\rm c}/u_{0}$ is controlled by the bottom slope.

\begin{figure}
  \includegraphics[scale=1]{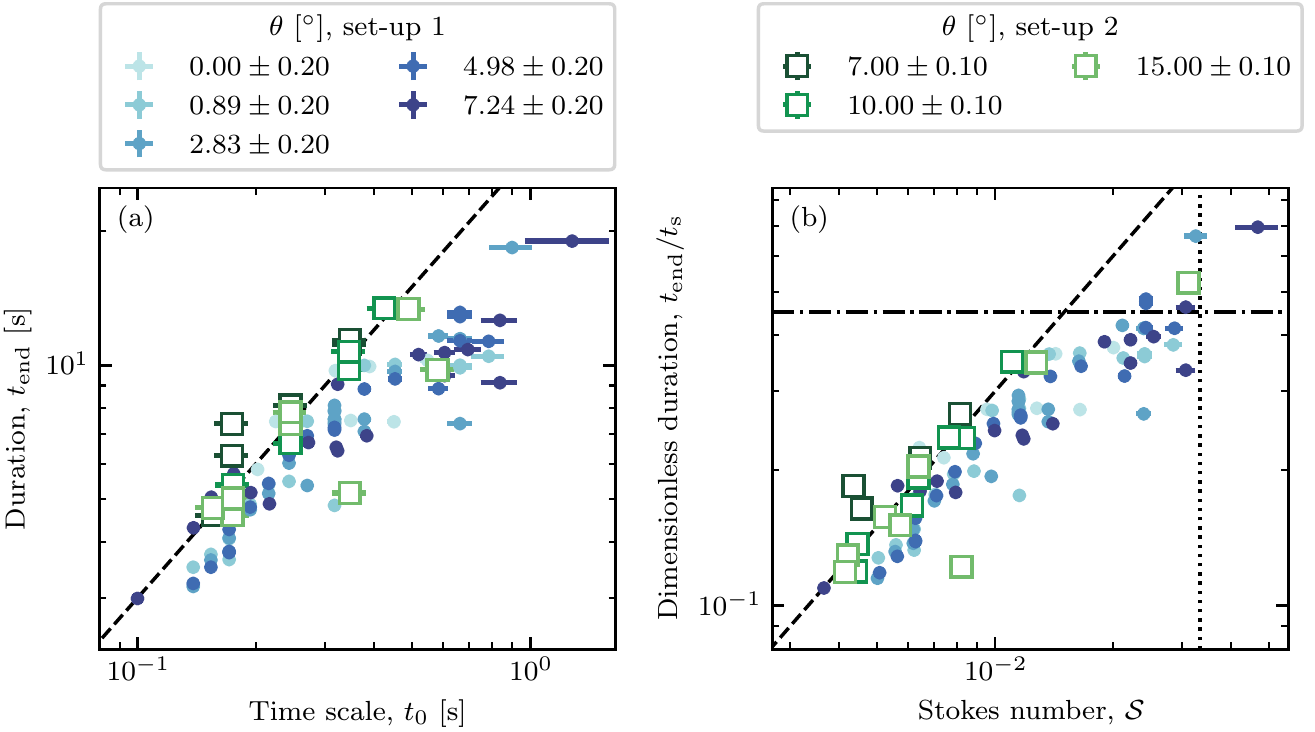}
  \caption{(a) Duration of the constant-velocity regime as a function of the current characteristic time scale, for various bottom slopes in the two experimental set-ups. (b) Same as (a), but with rescaling both axes by a settling time scale $t_{\rm s} = h_{0}/u_{\rm s}$. In both subplots, the black dashed line indicates $t_{\rm end} = 30 t_{0}$. On (d), the vertical dotted line indicates the limit $\mathcal{S} \simeq 0.033$ (see figure~\ref{fig:fig3}(b) and section~\ref{sec:settling_velocity}), and the horizontal dash-dotted line the limit $t_{\rm end} = 0.45t_{\rm s}$. Here, the settling velocity is $u_{\rm s} = 0.74~\textrm{cm}~\textrm{s}^{-1}$.}
  \label{fig:figure_tend_slope}
\end{figure}

\section{Current morphology during the slumping regime}
\label{sec:av_shape}

\begin{figure}
  \includegraphics[scale=1]{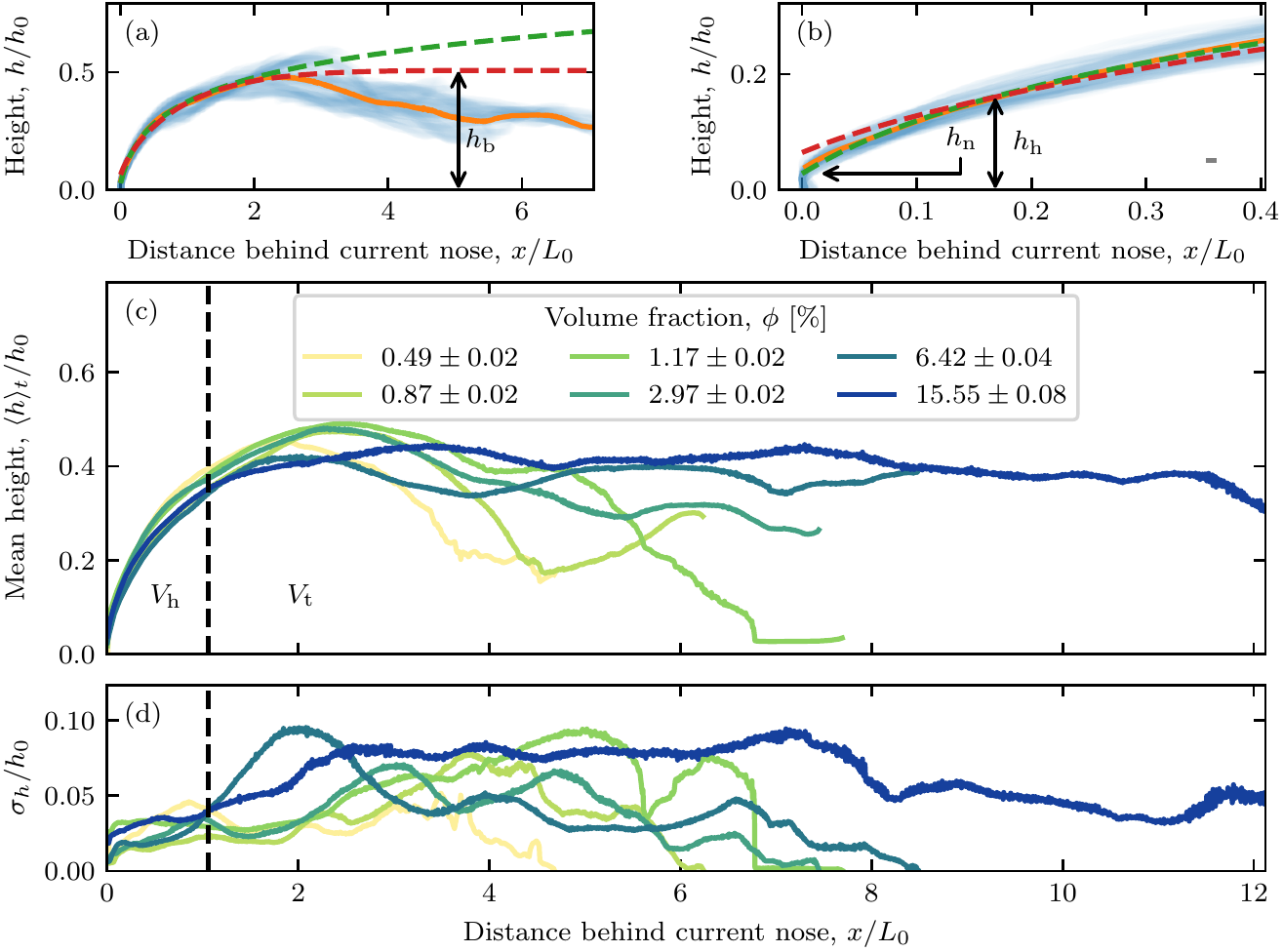}
  \caption{(a) Current shape for an experiment with an initial volume fraction $\phi=3~\%$. Blue lines: all shapes during the constant-velocity regime superimposed with transparency. Orange line: temporal average shape. Red dashed line: fit of Benjamin's current shape. Green dashed line: fit of logarithmic shape \eqref{eq:logshape}. (b) Zoom of (a) on the first centimetres. The grey rectangle indicates the camera pixel size. (c) Average shapes during the constant-velocity regime for various initial volume fractions. (d) The standard deviation corresponding to the shapes in (c). In (c) and (d), the black dashed lines separate the current head from its body. Not all experiments are shown, for the sake of clarity. In this figure, grains are silica sand with $d \sim 120~\mu\textup{m}$, and the bottom slope is $\theta=7^\circ$.}
  \label{fig:fig4}
\end{figure}

In this section, we focus on the current morphology. During the constant-velocity regime, the current shape is found to be defined by an average shape in the frame of the current nose (blue and orange curves in figure~\ref{fig:fig4}(a, b)). Fluctuations around this average profile can be quantified by the standard deviation as shown in figure~\ref{fig:fig4}(d).

\subsection{Morphometrics}
The quantitative characterization of the current shape has always been a challenge in the literature, aiming, for example, at the extraction of a current characteristic height. When velocity or density/concentration profiles are accessible, studies have used a height weighted by buoyancy \citep{shin2004, marino2005, Cantero2007, sher2015} or kinetic energy profiles \citep[e.g.][]{islam2010, stagnaro2014}. When a single contour is available, the height of the trailing current behind the head has been used widely, provided that it is well defined \citep[e.g.][]{simpson1980, bonnecaze1993, lowe2005, Chowdhury2011}.

As shown in figure~\ref{fig:fig4}(c), the shape of the observed currents spans from a single head (low volume fractions) to a continuous current with no distinguishable head lobe (highest volume fractions). The same qualitative variation is observed between low and high settling velocities, or for saline homogeneous currents between low and high excess density (not shown here).

In order to encompass all these morphologies, we use the following approach. First, we fit the theoretical shape of a steady current calculated by \citet{benjamin1968}, to which we add a free vertical shift to account for the nose (foremost point of the head) height induced by bottom friction (red dashed lines in figures~\ref{fig:fig4}(a,b)). This allows us to extract a current height $h_{\rm b}$, as well as the current nose height $h_{\rm n}$. While Benjamin's shape accounts for the large-scale behaviour of the current's head, it does not reproduce well the curvature close to the nose (dashed red line in figure~\ref{fig:fig4}(b)), and therefore leads to poor estimations of $h_{\rm n}$. However, we noticed that close to the nose, the current head is well approximated by a portion of a logarithm (see green dashed line in figures~\ref{fig:fig4}(a,b)):
\begin{equation}
  \label{eq:logshape}
  h(x) = h_{\rm h}\log\left(\frac{x + \delta}{x_{\rm c}}\right),
\end{equation}
where $\delta$ is a shift parameter, found to be almost constant for all currents, and therefore fixed to $1.4~\textrm{cm}$ ($0.14 L_{0}$). Here, $h_{\rm h}$ gives a characteristic head height representing its geometry, and $h(0) \equiv h_{\rm n}$ is the nose height.

Finally, we also noticed that the average current shape can be split into two parts (figures~\ref{fig:fig4}(c,d)). Close to the nose, the head presents relatively small variations during the current propagation (figure~\ref{fig:fig4}(d)), and is also rather invariant with respect to the volume fraction (see figure~\ref{fig:fig4}(c)), but also to the bottom slope and the settling velocity. On the contrary, the tail presents the largest temporal fluctuations induced by shear instabilities, and its morphology depends largely on the volume fraction and the settling velocity (see section~\ref{sec:results_morpho} for further discussion). Such observation suggests the spatial development of a Kelvin-Helmholtz instability along the interface of the current, from the head (at which its amplitude remains small) towards the tail of the current.
Accordingly, the transition between head and tail is defined as a change in the standard deviation, found to increase beyond a distance $L_{0}$ behind the current nose (black dashed line in figure~\ref{fig:fig4}(d)). The volume of the current head (per unit of width), $V_{\rm h}$, is then calculated on the corresponding distance of one $L_{0}$ behind the current nose (black dashde line in figure~\ref{fig:fig4}(c)). The volume of the tail (per unit of width), $V_{\rm t}$, is calculated as the total volume minus the head volume $V_{\rm h}$.

\begin{figure}
  \includegraphics[scale=1]{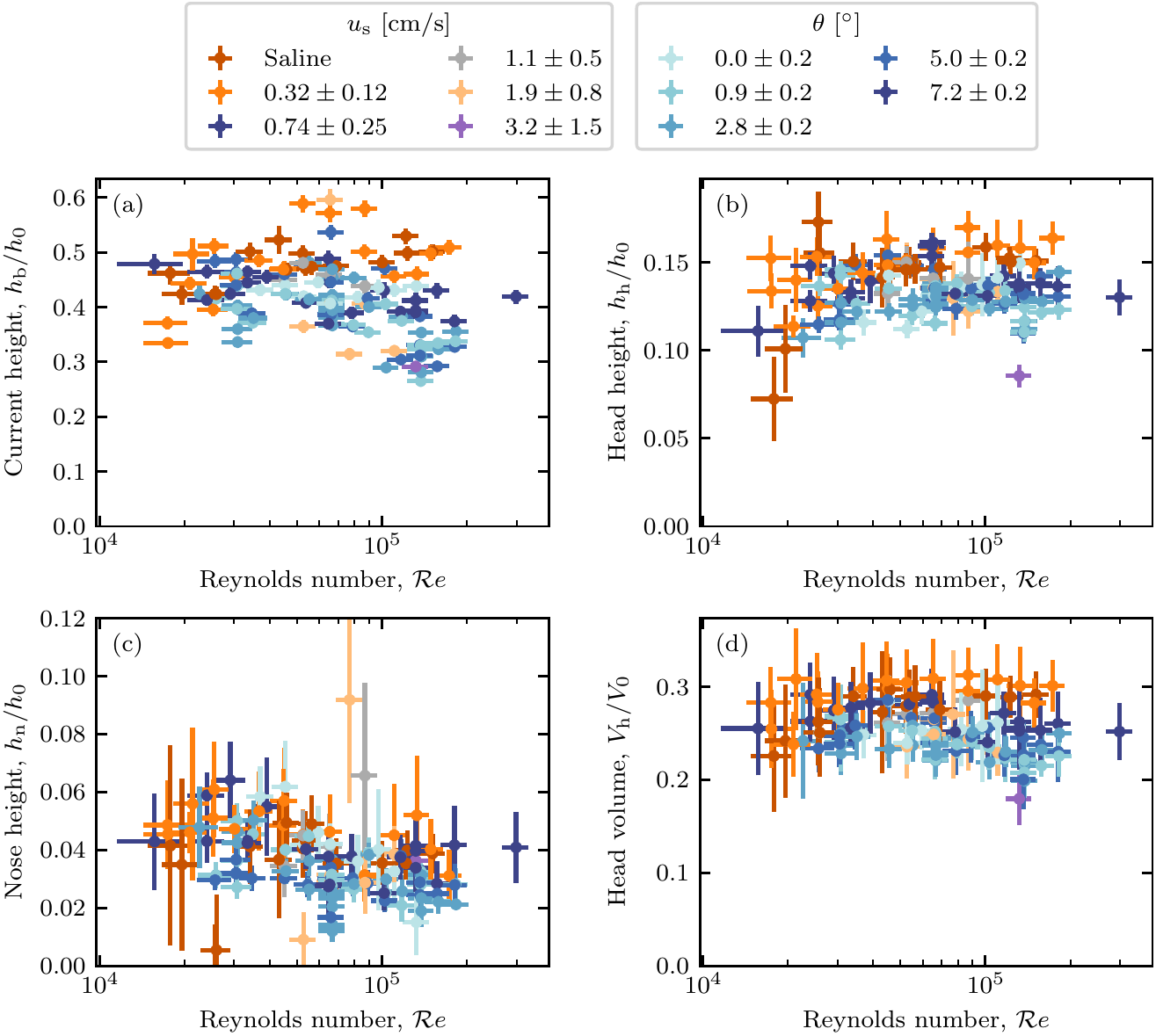}
  \caption{Average shape properties as a function of the bulk Reynolds number for various bottom slopes and settling velocities: (a) current height, (b) head height, (c) nose height, (d) current head volume.}
  \label{fig:fig5}
\end{figure}

\subsection{Results}
\label{sec:results_morpho}

\begin{figure}
  \includegraphics[scale=1]{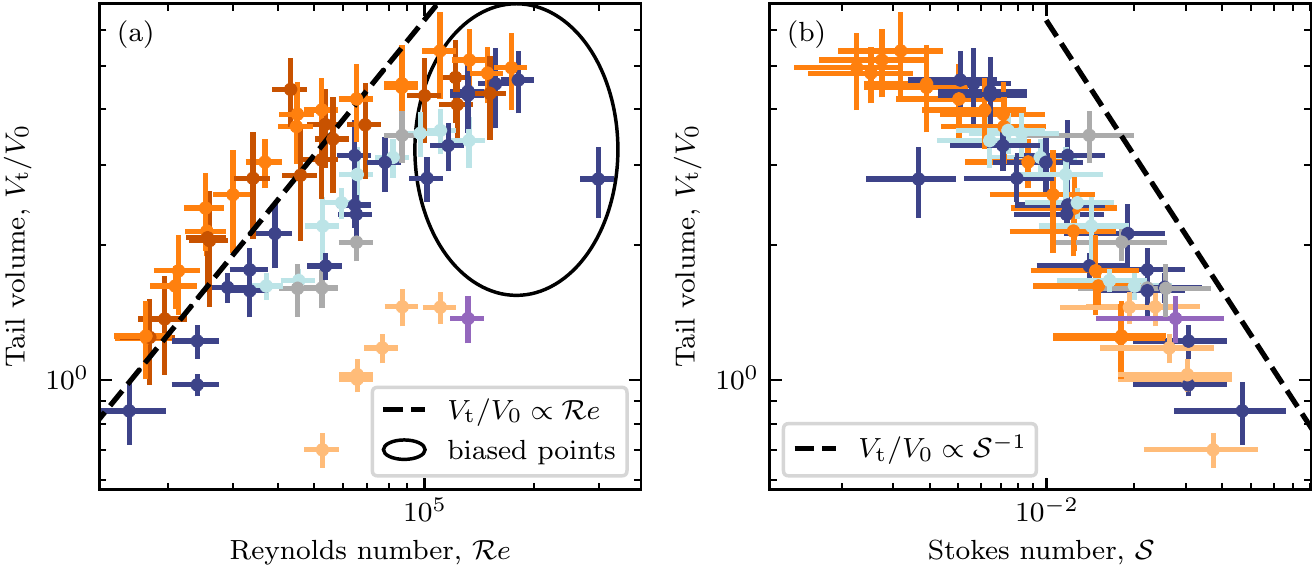}
  \caption{Average dimensionless current tail volume as a function of the bulk Reynolds number for various bottom slopes and settling velocities. Intermediate slopes are not shown to better highlight the effect of the settling velocity. In (a), biased points, typically for $\mathcal{R}e > 10^{5}$, represent runs for which we have little confidence in the tail volume due to the lock opening (see section~\ref{sec:results_entrainment} and Appendix~\ref{sec:appendix_door_opening} for further discussion). They are removed in (b). The legend for the colours is the same as in figure~\ref{fig:fig5}.}
  \label{fig:fig6}
\end{figure}

The characteristic quantities $h_{\rm b}$, $h_{\rm h}$, $h_{\rm n}$, $V_{\rm h}$ and $V_{\rm t}$ are shown in figure~\ref{fig:fig5} figure~\ref{fig:fig6} for all experimental runs that exhibit a constant-velocity regime (see section~\ref{sec:existence_regime}). A key observation is that all parameters linked to the current head morphology ($h_{\rm b}$, $h_{\rm h}$, $h_{\rm n}$, $V_{\rm h}$) are found to be independent of all experimental parameters, excess density (i.e. $\mathcal{R}e$), settling velocity (i.e. $\mathcal{S}$) and bottom slope (see figure~\ref{fig:fig5}). On the other hand, the volume of the tail, $V_{\rm t}$, is found to increase with the excess density (i.e. $\mathcal{R}e$) and decrease with the settling velocity (i.e. $\mathcal{S}$) (see figure~\ref{fig:fig6}).

\subsubsection{Current height}

As shown in figure~\ref{fig:fig5}(a), the average current height $h_{\rm b}$ is $\simeq 0.4~h_{0}$, in agreement with previous studies \citep{shin2004, sher2015}. All experimental points also lie within the predictions of Benjamin and single-layer shallow-water models, $h_{\rm b} = 0.5~h_{0}$, and two-layer shallow-water models, $h_{\rm b} = 0.35~h_{0}$~\citep{benjamin1968, ungarish2007, ungarish2011}. Note that a slight decrease can be observed for large excess densities, corresponding to large volume fractions and Reynolds numbers. This could result from non-Boussinesq effects, although they are predicted to be insignificant in the density ratios range of our experiments by shallow-water models \citep{ungarish2007, ungarish2011}.

Likewise, we obtain approximately constant nose and head heights $h_{\rm h} \simeq 0.13~h_{0}$ and $h_{\rm n} \simeq 0.04~h_{0}$ (figures~\ref{fig:fig5}(b, c)). Note that here, $h_{\rm n}/h_{\rm b} \simeq 0.1$, similar to previous measurements available in the literature for the same range of Reynolds numbers performed on saline homogeneous density currents (see figure 9 of \citet{hartel2000}, and corresponding measurements of \citet{barr1963}, \citet{keulegan1957} and others).

Despite the dispersion in our data, it also seems that saline homogeneous currents, and turbidity currents with the smallest settling velocity have in general higher heights than turbidity currents with larger settling velocities (see figures~\ref{fig:fig5}(a--c)). This could result from a less dilute interface induced by larger settling velocities, but investigating this requires additional dedicated experiments. Finally, the influence of the bottom slope remains negligible on the current height in the range $[0^\circ, 7^\circ]$.

\subsubsection{Current volume}

While the current head volume is constant, $V_{\rm h} \simeq 0.25 V_{\rm 0}$ (see~figure~\ref{fig:fig5}(d)), the tail volume increases with the Reynolds number (see~figure~\ref{fig:fig6}(a)). For saline currents and the smallest settling velocity, this increase is rather linear. However, increasing the settling velocity leads to smaller values of $V_{\rm t}/V_{\rm 0}$, which may correspond to a different slope of this relationship and/or to a different power law. A good collapse is obtained by plotting the experimental data as a function of the Stokes number, for which we find $V_{\rm t}/V_{\rm 0} \propto \mathcal{S}^{-1}$ (see~figure~\ref{fig:fig6}(b)).
The volume increase cannot be driven solely by the Stokes number, as this would imply that saline gravity currents, for which $\mathcal{S} = 0$, would have a constant $V_{\rm t}/V_{\rm 0}$ value, which is not the case. This means that $V_{\rm t}/V_{\rm 0}$ depends on both $\mathcal{R}e$ and $\mathcal{S}$.

While most currents have $V_{\rm t}/V_{\rm 0} \geq 1$, suggesting the presence of water entrainment, currents for $\mathcal{S} > 0.03$ have $V_{\rm t}/V_{\rm 0} \leq 1$, suggesting the dominance of particle settling. Then the dependence of $V_{\rm t}/V_{\rm 0} \geq 1$ with $(\mathcal{R}e,\mathcal{S})$ has to be related to entrainment, which is discussed in section~\ref{sec:water_entrainment}.


\section{Water entrainment}
\label{sec:water_entrainment}

\subsection{Parametrization and hypotheses}
\label{sec:parametrization_E}

Here, we consider a fixed observation window starting at the lock gate and ending at the end of the illuminated area (nearly the distance $L_1$ in figure~\ref{fig:fig1}, i.e. excluding the initial reservoir). In this zone, the continuity equation for the current volume $V$ (per unit of width) can be written as
\begin{equation}
  \label{eq:continuity}
  \frac{\textrm{d} V}{\textrm{d} t} = Q_{e} - Q_{s} + Q_{\textrm{in}},
\end{equation}
where $Q_{e}$ and $Q_{s}$ are fluxes induced by water entrainment and particle settling, respectively. As the observation window does not take into account what is inside the initial reservoir, an input flux $Q_{\textrm{in}}$ must be taken into account as long as part of the suspension is transferred from the reservoir to the current.

The entrainment flux can be written as the quantity of water passing through the interfacial line between the current and the ambient, $\Gamma$, at velocity $w_{e}$:
\begin{equation}
  Q_{e} = w_{e}\Gamma,
\end{equation}
where $w_{e} = E u_{\rm c}$ is the entrainment velocity \citep{jacobson2014}, and $E$ is the entrainment coefficient.

\begin{figure}
  \includegraphics[scale=1]{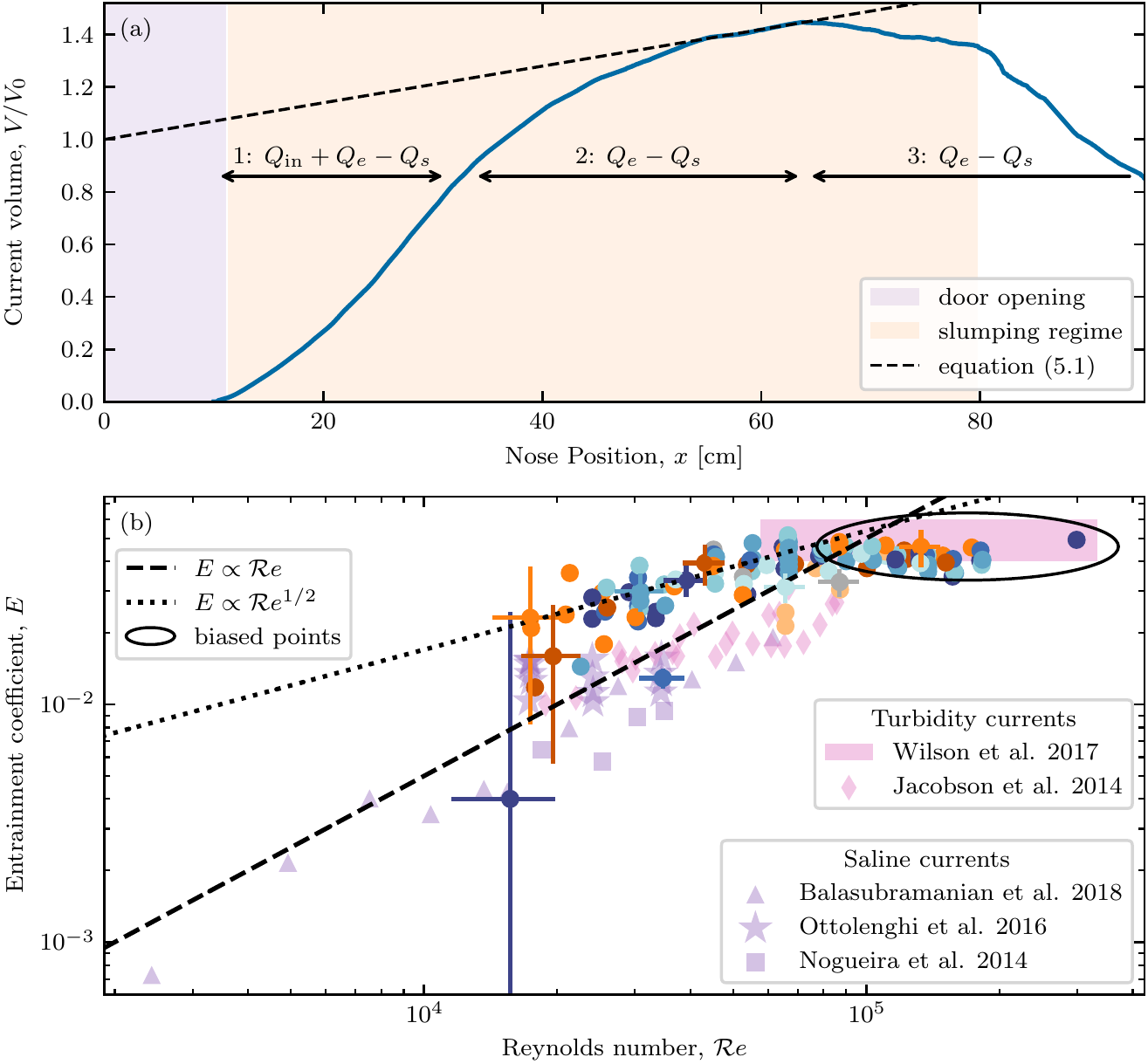}
  \caption{(a) Current volume as a function of nose position for $\theta = 7.2^\circ$, $u_{\rm s} = 0.32~\textrm{cm}~\textrm{s}^{-1}$ and $\phi = 0.24~\%$. The double arrows are indicative of the three different phases of the evolution of the measured current volume (see section~\ref{sec:parametrization_E} for details). (b) Water entrainment coefficient as a function of the bulk Reynolds number. Biased points, typically for $\mathcal{R}e > 10^{5}$, represent runs for which we have little confidence in the calculated entrainment due to the lock opening (see section~\ref{sec:results_entrainment} and Appendix~\ref{sec:appendix_door_opening} for further discussion). Not all error bars are shown, for the sake of clarity. The legend for the coloured circles is the same as in figure~\ref{fig:fig5}.}
  \label{fig:fig7}
\end{figure}

As shown in figure~\ref{fig:fig7}(a), the temporal evolution of the current volume, as modelled by \eqref{eq:continuity}, can be split into three phases.
Just after the lock opens, the volume increases due to the inflow $Q_{\rm in}$ at the upstream boundary (lock gate) induced by the column collapse (phase~1: $Q_{\rm in}>0, \, Q_{\rm e} \gg Q_{\rm s}$). After the reservoir has emptied, only entrainment and settling remain, during which the increase of the current volume becomes slower (phase~2: $Q_{\rm in}=0, \, Q_{\rm e} \gg Q_{\rm s}$). As the current increases its volume (entrainment) and loses some particles (settling), it dilutes up to a point where it gradually passes below the detection threshold chosen for the contour extraction (see figures~\ref{fig:fig1}(c--h)). At this point, the current volume starts to decrease, and \eqref{eq:continuity} is not longer applicable (phase~3: $Q_{\rm in}=0, \, Q_{\rm e} \ll Q_{\rm s}$).
The ubiquitous presence of settling, combined with the observed diversity of cases, makes it difficult to distinguish whether the volume increase is due solely to entrainment. Therefore, we compute a bulk entrainment parameter, by considering the volume difference between the maximum volume observed and the initial volume in the reservoir. We assume that at this time, the reservoir has emptied completely, and that the current velocity is still large enough to neglect settling processes. This assumption seems relevant as this time remains within the slumping regime, during which the current dynamics is not impacted by settling (see section~\ref{sec:slumping_behaviour}).
Following previous studies \citep{Cenedese2008, nogueira2014, jacobson2014, wilson2017}, the entrainment coefficient therefore reads
\begin{equation}
  E = \frac{1}{u_{\rm c}}\frac{\textrm{d} V}{\textrm{d} t}\frac{1}{\Gamma} = \frac{\textrm{d} V}{\textrm{d} x}\frac{1}{\Gamma}.
\end{equation}
Following the mentioned literature, the interfacial length $\Gamma$ is also taken at the time when the current volume reaches its maximum.

\subsection{Bulk entrainment coefficient}
\label{sec:results_entrainment}

The resulting entrainment coefficients are shown in figure~\ref{fig:fig7}(b) as a function of the Reynolds number, along with data from previous studies. Disregarding biased points (see below, and Appendix~\ref{sec:appendix_door_opening}), an increasing trend with the Reynolds number is visible. On the other hand, no clear impact of the bottom slope or the settling velocity (or here, equivalently, particle size) is found in the studied range. This is not surprising for the bottom slope, as \citet{Cenedese2008} have shown that slopes larger than $25^\circ$ are needed to increase water entrainment significantly.
Note that error bars in figure~\ref{fig:fig7}(b) for $\mathcal{R}e < 2~10^{4}$ are large because the volume variations induced by entrainment and by fluctuations during the propagation become of the same order.
Moreover, at large Reynolds numbers (typically $\mathcal{R}e > 10^{5}$), entrainment saturates to a constant value. However, we attribute this to bias induced when releasing the initial reservoir. For the corresponding runs, the release velocity is too small compared to the current velocity. This results in the mixing of a significant portion of the reservoir with the ambient fluid brought back by the overlying backflow (see Appendix~\ref{sec:appendix_door_opening} for further description). As such, these currents are still fed by an input flux, even though the slumping regime, controlled by the current head properties (formed before the opening-induced mixing occurs), is over. This results in an erroneously measured constant maximum volume.


Overall, as shown by figure~\ref{fig:fig7}(b), our data agree well with previous experimental studies on both saline \citep{nogueira2014, Ottolenghi2016, balasubramanian2018} and turbidity currents \citep{jacobson2014, wilson2017}, suggesting a dominant linear correlation between entrainment and $\mathcal{R}e$. Surprisingly, \citet{wilson2017} found constant entrainment values matching the saturation induced by the bias of our data at $\mathcal{R}e > 10^{5}$. Note that the data of \citet{balasubramanian2018} have been obtained by a direct method based on buoyancy fluxes, which validates further the entrainment parametrization used in this and other studies.
Despite the dispersion within each dataset, we find slightly larger entrainment coefficients. Note, however, that the absolute value of our results depends on the chosen threshold for the current contour extraction, which can lead to a volume variation corresponding to a vertical downward shift of $E$ of the order of $10^{-2}$ in our data.

Interestingly, a similar trend ($E$ increasing with $\mathcal{R}e$, followed by a saturation) has been found in laboratory and field data on constant-inflow turbidity and gravity currents in a rotating frame~\citep{Cenedese2010, Wells2010}. Note that the parametrization of \citet{Cenedese2010} tends to $E \propto \mathcal{R}e^{1/2}$ for small Reynolds numbers, a power law that could also match the trend of the data acquired in this study (see circles in figure~\ref{fig:fig7}). 
However, it should be noted that our dataset is shifted towards larger values of the Reynolds numbers, i.e. the increase from $E \simeq 10^{-3}$ to $E \simeq 10^{-2}$ is found to occur at $\mathcal{R}e \in [10^{2}, 10^{3}]$ for $\mathcal{F}r \approx 0.5$ in \citet{Cenedese2008}. Hence the corresponding $\mathcal{R}e$ value above which $E$ becomes independent of $\mathcal{R}e$ is also much larger. This suggests that the dependency of $E$ on $\mathcal{R}e$  could thus be of relevance for field scale gravity/turbidity currents. In any case, further work remains to be done to match the datasets on constant volume and constant inflow gravity/turbidity currents.

\section{Conclusion}

In the present study, we investigate the slumping regime, characterized as the constant front velocity regime, of lock-release turbidity currents using an experimental approach. In particular, we explore systematically the influence of volume fraction, bottom slope and particle settling velocity, which remains relatively sparse and scattered in the existing literature. For that purpose, we define the associated independent dimensionless parameters as the Reynolds number $\mathcal{R}e$, the Stokes number $\mathcal{S}$ and the slope $\theta$. Direct comparison is also made with saline homogeneous gravity currents for which $\mathcal{S}\equiv 0$.

In the explored parameter range, saline and turbidity currents exhibit a constant-velocity regime, i.e. a slumping regime, only if $\mathcal{S} \lesssim 0.033$. We then show that each parameter dominantly controls one specific property of the slumping regime.
\begin{itemize}
  \item The dimensionless current velocities $\mathcal{F}r$ (i.e. Froude numbers) increase with $\theta$, while being independent of $\mathcal{R}e$ and $\mathcal{S}$. A relevant energetic balance during this transient regime, including along-slope weight and friction, is found here to provide the relevant slope effect as $\mathcal{F}r(\theta) = \mathcal{F}r_0 \sqrt{\cos\theta + C\sin\theta}$. The values of $\mathcal{F}r_0$ and $C$ appear to depend on the experimental device used, thus will require dedicated attention in future studies.

  \item The duration of the slumping regime $t_{\rm end}$ depends on $\mathcal{S}$. For $\mathcal{S} \lesssim 0.01$, $t_{\rm end} \simeq 30 t_{0}$ as for saline homogeneous currents. As $\mathcal{S}$ increases, the regime duration decreases, up to being fully controlled by settling, i.e. $t_{\rm end} \simeq 0.45 h_{0}/u_{\rm s}$, or equivalently, $t_{\rm end}/t_{0} \simeq 0.45 \mathcal{S}^{-1}$. For $\mathcal{S} \gtrsim 0.033$, the slumping regime disappears.

  \item Entrainment during the slumping regime can be characterized by a time-independent entrainment coefficient $E$. For the parameter range covered here, $E$ is found to increase rather linearly with $\mathcal{R}e$, while being independent of $\mathcal{S}$ and $\theta$.
\end{itemize}
Interestingly, the morphology of the current head is found to be independent of $\mathcal{R}e$, $\mathcal{S}$ and $\theta$. Above a sublayer induced by bottom friction, the head shape is well approximated by the theoretical shape of Benjamin's current, with $h_{\rm b}/h_{0} \simeq 0.4$. However, close to the nose, the head is found to be further curved downwards, presumably due to the influence of bottom friction, and better approximated by a portion of a logarithm, with $h_{\rm h}/h_{0} \simeq 0.1$ and $h_{\rm n}/h_{0} \simeq 0.04$.

Overall, this work supports the modelling of turbidity currents as an average fluid of equivalent density $\rho_{0}$ as long as $\mathcal{S} \lesssim 0.01$. When $\mathcal{S} \gtrsim 0.01$, the current dynamics differ due to the presence of the settling particles, which in this case would require dedicated modelling, similarly to non-Newtonian effects in the fluid rheology for large particle volume fractions \citep{Chowdhury2011, jacobson2014}.
Nevertheless, in order to completely unravel the origin of the slope effect on the Froude number, further work on the early transient regime in the case of an inclined bottom is still required, especially concerning the influence of the lock geometry.

\backsection[Acknowledgements]{We thank Jean-Dominique Barron (IMFT) and Sébastien Cazin (IMFT) for their support in carrying out the experiments on setup-up 1, as well as Sylvain Dauge for the design of setup-up 2. Finally, we would like to acknowledge the contributors of the open-source Python libraries, including Matplotlib \citep{Hunter2007}, NumPy \citep{Harris2020} and SciPy \citep{Virtanen2020}, which provide an incredibly efficient ecosystem allowing scientific research in Python.}

\backsection[Funding]{We acknowledge financial support from the French National Research Agency Grants, ANR-19-CE30-0041/PALAGRAM.}

\backsection[Declaration of interests]{The authors report no conflict of interest.}

\backsection[Data availability statement]{The data that support the findings of this study are openly available in Zenodo at https://doi.org/10.5281/zenodo.7487189.}

\backsection[Author ORCID]{C. Gadal, https://orcid.org/0000-0002-2173-5837; M. J. Mercier, https://orcid.org/0000-0001-9965-3316; M. Rastello, https://orcid.org/0000-0002-4457-1433; L. Lacaze, https://orcid.org/0000-0002-4945-7445}

\appendix

\section{Grain properties}
\label{sec:grain_properties}

\subsection{Grain size distributions}

The particle size distributions are obtained by taking pictures of the grains using a microscope. The resulting images are segmented using the CellPose algorithm \citep{Pachitariu2022}, leading to a collection of planar shapes for each particle type. For each shape, three different diameters are calculated: an average diameter assuming a circular shape, and the major and minor axis of the ellipse that has the same normalized second central moments as the selected shape.

\begin{figure}
  \includegraphics[scale=1]{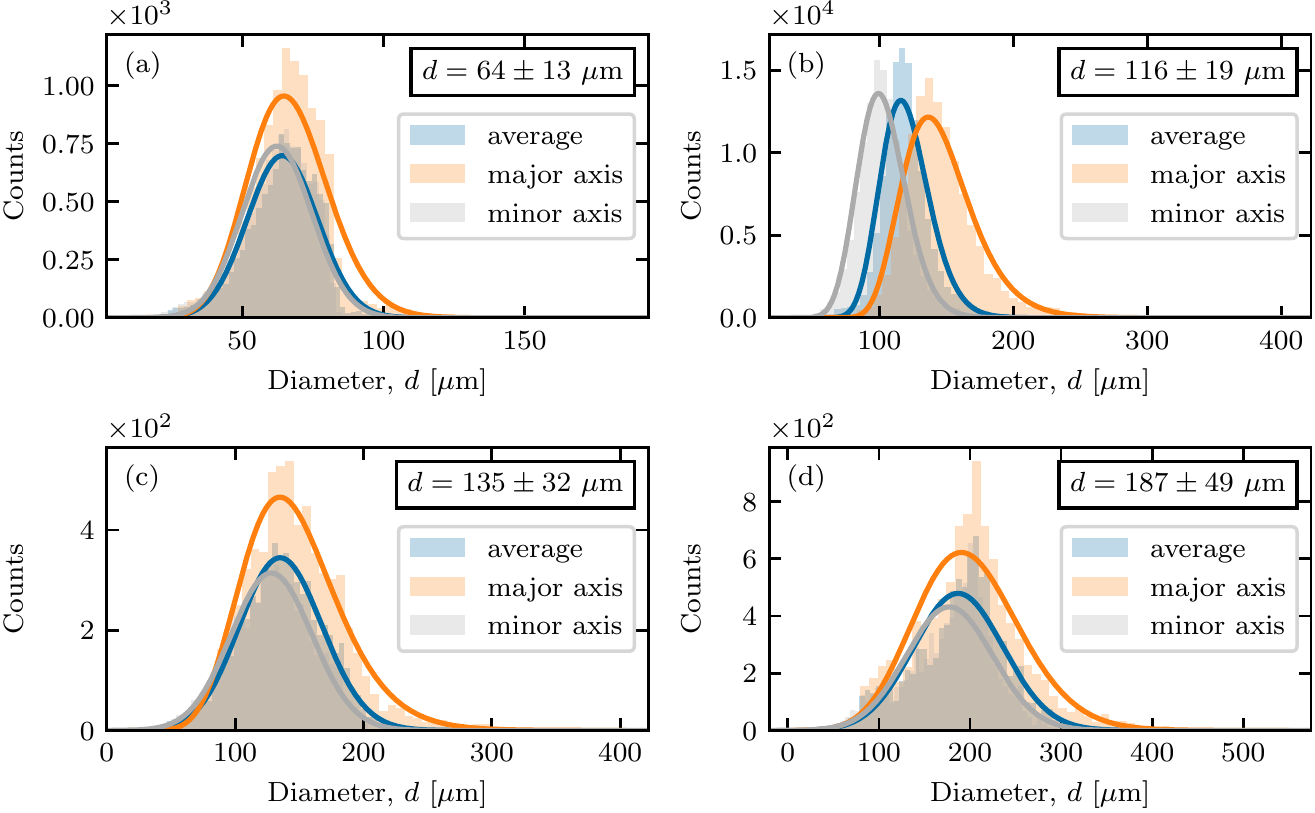}
  \caption{Grain size distributions for the particles used in the paper: (a) Silibeads 40--70 $\mu\textrm{m}$, (b) sand 120 $\mu\textrm{m}$, (c) Silibeads 100--200 $\mu\textrm{m}$, (d) Silibeads 150--250 $\mu\textrm{m}$. The solid lines are fits of log-normal distributions, and the modal value of the average diameter distribution is shown at the upper right of each plot.}
  \label{fig:grain_distributions}
\end{figure}

The resulting distributions are shown in figure~\ref{fig:grain_distributions}. For the glass beads, all three diameters exhibit similar distributions with matching modes. For the silica sand, the average diameter is between the minor and major axes of the corresponding ellipse. Note that the measurements for the Silibeads 200--300 $\mu\textrm{m}$ are lacking, due to problems with the microscope. However, for the other glass beads, the measured distributions are in fairly good agreement with the range given by the manufacturer. Therefore, for the Silibeads 200--300 $\mu\textrm{m}$, we take $d = 250~\mu\textrm{m}$.

\subsection{Settling velocity}

The particle settling velocity is calculated from the equilibrium between buoyancy,
\begin{equation}
  f_{\rm g} = \frac{1}{6}\pi(\rho_{\rm p} - \rho_{\rm f}) g d^{3},
\end{equation}
and the drag force,
\begin{equation}
  f_{\rm d} = \frac{1}{8}\rho_{\rm f}u_{\rm s}^{2}\pi d^{2} C_{\rm D},
\end{equation}
where $C_{\rm D}$ is a drag coefficient, a function of the particle Reynolds number $\mathcal{R}_{\rm p} = u_{\rm s} d/\nu$ and therefore of the settling velocity. Various forms of the drag coefficient can be found in the literature~\citep{van2008}. Here, we follow the approach of \citet{Camenen2007} by writing the drag coefficient in the form
\begin{equation}
  C_{\rm D} = \left[\left(\frac{A}{\mathcal{R}_{\rm p}}\right)^{1/m} + B^{1/m}\right]^m,
\end{equation}
where $A$ and $B$ are two constants that depend on the particle shape. Balancing the two forces thus leads to the following expression for the settling velocity:
\begin{equation}
  \label{eq:settling_velocity}
  \frac{\nu}{d} u_{\rm s} = \left[\sqrt{\frac{1}{4}\left(\frac{A}{B}\right)^{2/m} + \left(\frac{4}{3}\frac{d_{*}^{3}}{B}\right)^{1/m}}  - \frac{1}{2}\left(\frac{A}{B}\right)^{1/m}\right]^{m},
\end{equation}
where $d_{*} = ((s-1)g/\nu^{2})^{2/3}d$ is a dimensionless particle diameter, and $s = \rho_{\rm p}/\rho_{\rm f}$. Following the empirical calibration by \citet{Camenen2007}, we use $A = 24$, $B = 0.4$ and $m = 1.92$, which corresponds to spherical particles.

\begin{figure}
  \includegraphics[scale=1]{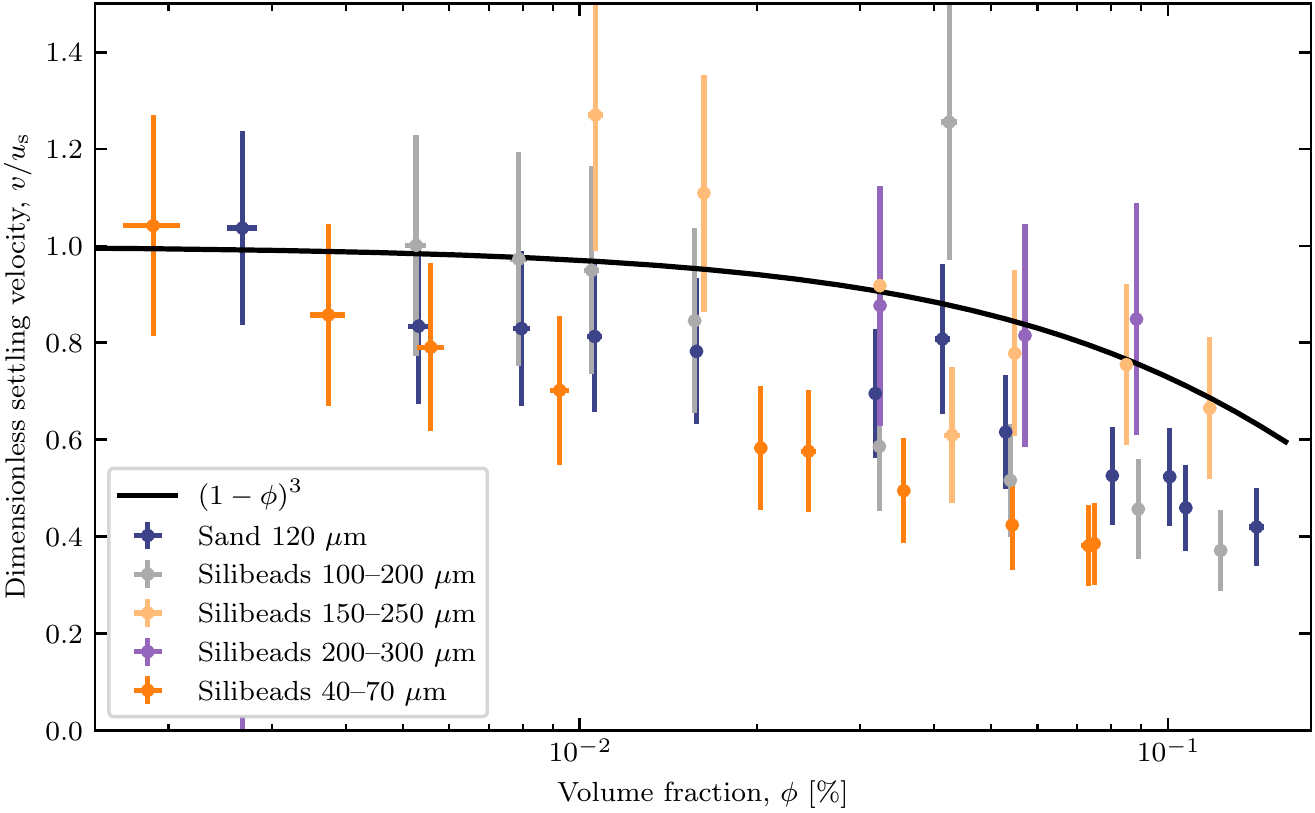}
  \caption{Dimensionless measured particle velocity as a function of the particle volume fraction. Note that the error bars essentially come from the uncertainty in the calculation of $u_{\rm s}$ from \eqref{eq:settling_velocity}, inherited from the parameter uncertainties (grain size, water viscosity, densities).}
  \label{fig:fig_settling_velocity}
\end{figure}

To check the calculated settling velocities, we use a simple experimental set-up in which we put the particles in suspension in a fluid column by stirring strongly, and then follow the front of the suspension as the particle sediments. As shown by figure~\ref{fig:fig_settling_velocity}, the calculated settling velocities match the experimental ones for dilute enough volume fractions. However, the measured settling velocity decreases with the volume fraction as observed previously in the literature~\citep{richardson1954}. Note that the observed decrease is faster than the typical correction in $(1 - \phi)^{1/3}$ proposed by \citet{richardson1954}, especially at low-volume fractions. According to \citet{diFelice1995}, the \citet{richardson1954} regime is reached only for volume fractions larger than $10~\%$. For more dilute suspensions, the decrease of the settling velocity with $\phi$ is stronger. Thus we leave out this complex dependence on particle volume fraction, and restrict ourselves to the settling velocities calculated using \eqref{eq:settling_velocity}.

\section{Entrainment induced by the door opening}
\label{sec:appendix_door_opening}

\begin{figure}
  \includegraphics[scale=1]{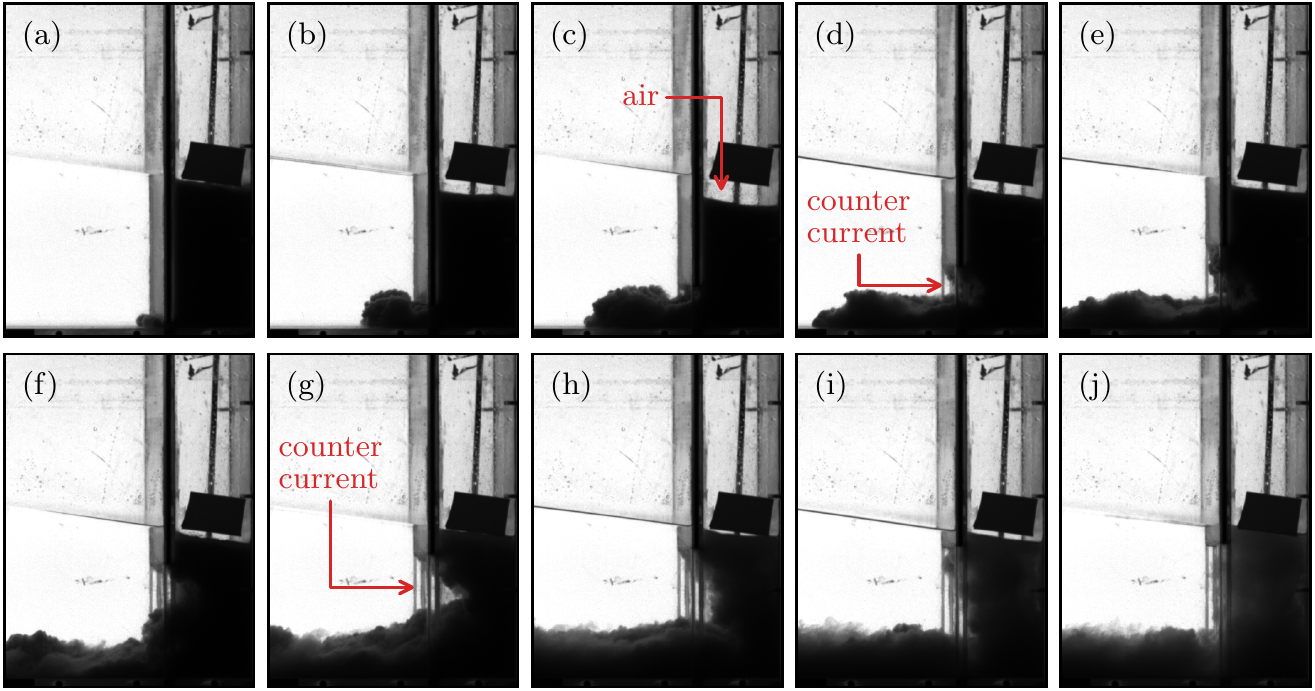}
  \caption{Close-up on the opening of the door for an experiment made with silica sand ($d \sim 120 \mu$m, $u_{\rm s} = 0.74~\textrm{cm}~\textrm{s}^{-1}$) for $\phi \sim 8~\%$.}
  \label{fig:figure_door_opening}
\end{figure}

In section~\ref{sec:results_entrainment}, we observed that the measured entrainment coefficient saturates to a constant value for $\mathcal{R}e > 10^{5}$. This is attributed to the opening of the lock gate, snapshots of which are shown in figure~\ref{fig:figure_door_opening}.

The first source of entrainment is induced at the beginning of the gate opening. As shown by figures~\ref{fig:figure_door_opening}(a--c), in the beginning, the tank empties as the suspension flows out of the bottom with no opportunity for the ambient fluid to create a counter-current at the top (the locked door is impermeable).
As soon as the gate has opened higher than the height of the current (figures~\ref{fig:figure_door_opening}(d, e)), the ambient fluid creates a counter-current just above the turbidity current, thus mixing the ambient fluid with the suspension and refilling the lock. Note that this first mechanism induces a dilution of approximately $\lesssim 10~\%$ of the suspension behind the lock (inferred from the reservoir volume to be filled in figure~\ref{fig:figure_door_opening}(c)).

A second source of entrainment occurs when the suspension column begins to collapse. As shown by figures~\ref{fig:figure_door_opening}(f--h), the column collapse begins at the level of the bottom of the door, not properly at the top of the lock. This creates an intrusion of ambient fluid inside the lock, surrounded at the top and bottom by the suspension (figure~\ref{fig:figure_door_opening}(g)). This unstable situation is resolved quickly by the collapse of the upper part of the suspension, which mixes with the ambient fluid below (figures~\ref{fig:figure_door_opening}(h, i)). The result, at the end of the gate opening, is a full lock of suspension at a smaller volume fraction than the initial one, although a large volume of suspension has already been released into the turbidity current.

For these runs, the reservoir then becomes much larger than its initial volume, causing the resulting currents to fill the entire length of the tank. The corresponding maximum current volumes are therefore constant, corresponding approximately to $h_{\rm h}L_{1}$, and so are the corresponding entrainment coefficients.
Note that the current head, which controls the current dynamics during the slumping regime, has the appropriate initial volume fraction since it forms before the second mechanism occurs.

\bibliographystyle{jfm}
\bibliography{bibliography}

\end{document}